\documentclass[aps,nofootinbib,showpacs,prd]{revtex4}
\usepackage{graphicx}

\usepackage{color}

\newcommand{\eps}{\varepsilon}

\newcommand{\AHEP}{Instituto de F\'{\i}sica Corpuscular --
  C.S.I.C./Universitat de Val{\`e}ncia \\
  Campus de Paterna, Apt 22085,
  E--46071 Val{\`e}ncia, Spain}
\newcommand{\Cinvestav}{Departamento de F\'{\i}sica, Centro de
  Investigaci{\'o}n y de Estudios Avanzados del IPN\\ Apdo. Postal
  14-740 07000 Mexico, DF, Mexico}

\begin{document}

\preprint{IFIC/11-09}

\title{Global constraints on muon-neutrino non-standard interactions}

\author{F. J. Escrihuela}\email{franesfe@alumni.uv.es}
\affiliation{\AHEP} 

\author{O. G. Miranda}\email{Omar.Miranda@fis.cinvestav.mx}
\affiliation{\Cinvestav}

\author{M. T\'ortola}\email{mariam@ific.uv.es}
\affiliation{\AHEP}

\author{J. W. F. Valle}\email{valle@ific.uv.es, URL: http://astroparticles.es/}
\affiliation{\AHEP}

\begin{abstract}
  
  The search for new interactions of neutrinos beyond those of the
  Standard Model may help to elucidate the mechanism responsible for
  neutrino masses. Here we combine existing accelerator neutrino data
  with restrictions coming from a recent atmospheric neutrino data
  analysis in order to lift parameter degeneracies and improve limits
  on new interactions of muon neutrinos with quarks.
  In particular we re-consider the results of the NuTeV experiment in
  view of a new evaluation of its systematic uncertainties.  We find
  that, although constraints for muon neutrinos are better than those
  applicable to tau or electron neutrinos, they lie at the few $\times
  10^{-2}$ level, not as strong as previously believed. We briefly
  discuss prospects for further improvement.

\end{abstract}


\maketitle

\section{Introduction}

The discovery of neutrino oscillations provides the only firm evidence
for new physics we currently have~\cite{Nakamura:2010zzi}, namely
neutrino mass. This constitutes the most important discovery in
particle physics in the last quarter century, implying that the
Standard Model, which correctly accounts for all other experimental
data in particle physics, needs revision~\cite{nunokawa:2007qh}.
Precision is the keyword in neutrino physics today, including solar
and atmospheric neutrino data analysis, as well as the determination
of neutrino masses and mixing parameters at reactor and long-baseline
accelerator
experiments~\cite{Schwetz:2011qt,schwetz:2008er,maltoni:2004ei}.
It is important not only to investigate the potential of current and
upcoming long baseline experiments to determine the neutrino
oscillation parameters, but also to probe for the possible existence
of non-standard neutrino interactions, NSI, for short. The latter are
expected in most models of neutrino mass generation, such as
seesaw-type schemes~\cite{schechter:1980gr} and will play a crucial
role since they will shed light on the scale characterizing this
so-far elusive mass generation mechanism.
The interest on probing for the existence of NSI has been growing over
recent years, thanks also to the increasing precision of upcoming
experiments~\cite{bandyopadhyay:2007kx,Huber:2009cw} and to the fact
that the current oscillation interpretation is not yet fully
robust~\cite{miranda:2004nb} so one needs to scrutinize the interplay
between oscillations and NSI in future
experiments~\cite{huber:2001de,huber:2002bi,huber:2001zw,Huber:2010dx}.

A wide class of non-standard neutrino interactions may be parametrized
at low energies by the effective Lagrangian
\begin{equation}
-{\cal L}^{eff}_{\rm NSI} = \sum_{\alpha\beta} 
\varepsilon_{\alpha \beta}^{fP}{2\sqrt2 G_F} (\bar{\nu}_\alpha \gamma_\rho L \nu_\beta)
( \bar {f} \gamma^\rho P f ) \,,
\label{eq:efflag}
\end{equation}
where $G_F$ is the Fermi constant and the parameters
$\varepsilon_{\alpha \beta}^{fP}$ characterize the strength of the
NSI. For simplicity, these are assumed to be real. The chiral
projectors $P$ denote $\{R,L=(1\pm\gamma^5)/2\}$, while $\alpha$ and
$\beta$ label the three neutrino flavors, $e$, $\mu$ and $\tau$ and
$f$ is a first generation charged SM fermion ($e, u$ or $d$).

Collider experiments produce well-controlled and clean muon neutrino
beams, with just a small component of electron neutrinos, while tau
neutrino beams are unavailable.
As a result one can expect that the muon neutrino NSI parameters,
$\varepsilon_{\mu \beta}^{fP}$ are better constrained in comparison
with other neutrino flavors. This is indeed the case for the
interaction with
electrons~\cite{Barranco:2007ej,Bolanos:2008km}. However, for the
interaction of neutrinos with quarks, although neutrino nucleon
scattering has a very long history since CDHS~\cite{Blondel:1989ev}
and CHARM~\cite{Allaby:1987vr}, the accuracy of these early
experiments was worse than that of the more recent NuTeV experiment,
which has measured the $\nu_\mu N$ interaction with a very high
accuracy~\cite{Zeller:2001hh}, reporting a discrepancy with the
Standard Model predictions.
While this may be interpreted as a hint of new physics, indicating at
face-value a potentially non-zero value for the NSI parameters
$\varepsilon_{\mu \alpha}^{fP}$~\cite{Davidson:2003ha}, uncertainties
coming from QCD corrections might have been
underestimated~\cite{Davidson:2001ji}.

Moreover, as expected, limits on muon neutrino NSI coupling
strengths~\cite{Davidson:2003ha} derived on the basis of 1-loop
dressing of the neutrino effective four-fermion vertex can not be
formulated rigorously. Indeed these are highly model dependent and,
even when a full analysis is performed, including all the necessary
diagrams required in order to obtain gauge-invariance, one does not
strongly constrain the flavor changing parameters
$\varepsilon^{qL,R}_{\mu\tau}$ and $\varepsilon^{qL,R}_{\mu
  e}$~\cite{Biggio:2009kv}.

Recently, two new determinations of the Standard Model electroweak
mixing parameter $\sin^2\theta_W$ from the NuTeV measurement have been
presented~\cite{Ball:2009mk, Bentz:2009yy}.
Given that the most stringent model--independent bounds on
$\eps_{\mu\alpha}$ come from the NuTeV data, it is necessary to
consider the effect of including these new corrections on the NuTeV
results.
Therefore, the main goal of this paper is to the review the current
status of the constraints for the $\nu_\mu$ non-standard neutrino
interactions in view of the larger uncertainties indicated by these
recent papers. 
In so-doing we will not only take into account the new results
mentioned above, but also combine the laboratory constraints with the
restrictions inferred from the analysis of atmospheric neutrino
data. In order to obtain these constraints we adopt as a simplifying
working hypotesis that all NSI parameters other than $\varepsilon_{\mu
  \alpha}^{qP}$ vanish.
We will see that restricting ourselves to a two-generation NSI global
analysis we obtain relatively stringent constraints on the NSI
interactions for muon neutrinos, at the few $\times 10^{-2}$ level,
thanks largely to the interplay of atmospheric data.

The paper is planned as follows.  In the next section we will make a
brief description of the neutrino nucleon scattering parameters that
are relevant for our analysis.  In
section~\ref{sec:constr-from-nutev}, we will discuss the NuTeV data
and their implications for the NSI parameters, while in
sec.~\ref{sec:constr-from-charm} we discuss the previous experiments
CHARM and CDHS.  In section~\ref{sec:comb-with-atmosph} we combine
this information with constraints coming from the analysis of
atmospheric neutrinos in order to obtain global constraints on the NSI
parameters.
Finally, in section~\ref{sec:concl-future-prosp} we will give our
conclusions and discuss the prospects for further improvement in the
determination of NSI parameters.

\section{$\theta_W$ measurements in neutrino--nucleon scattering
  experiments}
\label{sec:thet-meas-at}

Neutrino scattering experiments provide one of the most precise probes
of the weak neutral current, and have been often been used to measure
the electroweak mixing angle $\sin^2\theta_W$.  In particular, it has
been shown that experiments with an isoscalar target are particularly
convenient, since in this case the uncertainties due to unknown
corrections to the QCD parton model cancel to a large
extent~\cite{LlewellynSmith:1983ie}.
For an isoscalar target of up and down-type quarks, the largest
contributions to the neutral and charged current cross sections are
related by isospin invariance and their ratio can be written as:
\begin{equation}
  R^\nu = \frac{\sigma(\nu_\mu N\to\nu_\mu X)}{\sigma(\nu_\mu N \to \mu^- X)} = 
  (g^L_\mu)^2 + r(g^R_\mu)^2 
\label{eq:ratio-nu}
\end{equation}
\begin{equation}
  R^{\bar{\nu}} = \frac{\sigma({\bar{\nu}_\mu} N\to{\bar{\nu}_\mu} X)}
  {\sigma({\bar{\nu}_\mu} N \to \mu^+ X)} =
  (g^L_\mu)^2 + \frac{1}{r}(g^R_\mu)^2 
\label{eq:ratio-antinu}
\end{equation}
where one introduces the ratio 
\begin{equation}
r = \frac{\sigma({\bar{\nu}_\mu} N \to \mu^+ X)}{\sigma(\nu_\mu N\to\mu^- X)}
\end{equation}
and the coupling constants are defined as
\begin{equation}
(g^P_\mu)^2=(g^{uP}_\mu)^2+(g^{dP}_\mu)^2
\end{equation}
with P=L,R. In the presence of NSI of muon neutrinos with quarks,
these coupling constants are replaced by:
\begin{eqnarray}
\label{eq:gl-nsi} 
 ({\tilde{g}^L_{\mu}})^2 & =& (g^{uL}_{\mu}+\eps_{\mu\mu}^{uL})^2+(g^{dL}_{\mu}+\eps^{dL}_{\mu\mu})^2 
                     + \sum_{\alpha\neq\mu}|{\eps^{uL}_{\mu\alpha}}|^2 + 
                     \sum_{\alpha\neq \mu}|{\eps^{dL}_{\mu\alpha}}|^2\\
({\tilde{g}^R_{\mu}})^2 & =& (g^{uR}_{\mu}+\eps^{uR}_{\mu\mu})^2+(g^{dR}_{\mu}+\eps^{dR}_{\mu\mu})^2 
                     + \sum_{\alpha\neq\mu}|{\eps^{uR}_{\mu\alpha}}|^2 + 
                     \sum_{\alpha\neq \mu}|{\eps^{dR}_{\mu\alpha}}|^2 
\label{eq:gr-nsi}
\end{eqnarray}

The quantities $R^\nu$ and $R^{\bar{\nu}}$ have been measured in the
past at the CDHS~\cite{Blondel:1989ev} and CHARM
experiments~\cite{Allaby:1987vr}.  Another well-known observable for
the study of deep inelastic neutrino scattering on an isoscalar target
is the Paschos-Wolfenstein (PW) ratio defined as~\cite{Paschos:1972kj}
\begin{equation}
  R_{PW}  =   \frac{\sigma(\nu_\mu N \to \nu_\mu X) - \sigma(\bar\nu_\mu N
    \to \bar\nu_\mu X)} {\sigma(\nu_\mu N \to \mu^- X) -
    \sigma(\bar\nu_\mu N \to \mu^+ X)}  =
  \frac{R^\nu - r R^{\bar{\nu}} }{1-r} 
  =    (g^L_\mu)^2 - (g^R_\mu)^2 .
  \label{eq:PW}
\end{equation}
This ratio is particularly useful because it depends very weakly on
the hadronic structure of the nucleus target and it is largely
insensitive to the uncertainties resulting from charm production as
well as charm and strange sea distributions. However, the simultaneous
measurement of neutral current cross sections for neutrinos and
antineutrinos requires the use of separate neutrino and antineutrino
beams.
The NuTeV Collaboration makes use of this observable in order to
measure the electroweak mixing angle. Actually, they measure
experimentally the ratios $R_\nu$ and $R_{\bar{\nu}}$, shown in
Eqs.~(\ref{eq:ratio-nu}) and (\ref{eq:ratio-antinu}), which later
transform in the Paschos-Wolfenstein ratio $R_{PW}$ through their fit
procedure.

\section{Constraints from NuTeV  data}
\label{sec:constr-from-nutev}

The NuTeV collaboration used high statistics neutrino and antineutrino
beams to measure their neutral and charged current cross sections on
an iron target.
Using a statistical separation of the NC and CC event candidates,
based on the length of each event in the detector, NuTeV reported
experimental values for $R_\nu$ and $R_{\bar{\nu}}$, as mentioned in
the previous section. A numerical fit of these results provides a
measurement of the left and right--handed neutral couplings to the
light quarks~\cite{Zeller:2001hh}:
\begin{eqnarray}
(g_\mu^L)^2 & = & 0.30005\pm0.00137 \\
 (g_\mu^R)^2 & = & 0.03076\pm0.00110
\end{eqnarray}
Notice the discrepancy with respect to the Standard Model
expectations~\cite{Nakamura:2010zzi}:
\begin{equation}
(g_\mu^L)^2_{\text{SM}}  =  0.30399 \pm 0.00017 \quad
(g_\mu^R)^2_{\text{SM}} = 0.03001 \pm 0.00002\, ,
\end{equation}
due to the NuTeV preference for a lower effective left--handed
coupling, almost 3$\sigma$ away from the SM expectation.
Similarly, a value for the electroweak mixing angle in the on--shell
scheme was obtained from a numerical fit:
\begin{equation}
\sin^2\theta_W = 0.22773 \pm 0.00135^{stat} \pm 0.00093^{syst}
\end{equation}
which is $3\sigma$ away from  the value determined in global precision
electroweak fits~\cite{Nakamura:2010zzi}:
\begin{equation}
\sin^2\theta_W = 0.22292 \pm 0.00082
\end{equation}
\begin{figure*}
\begin{center}
\includegraphics[width=.45\textwidth,angle=0]{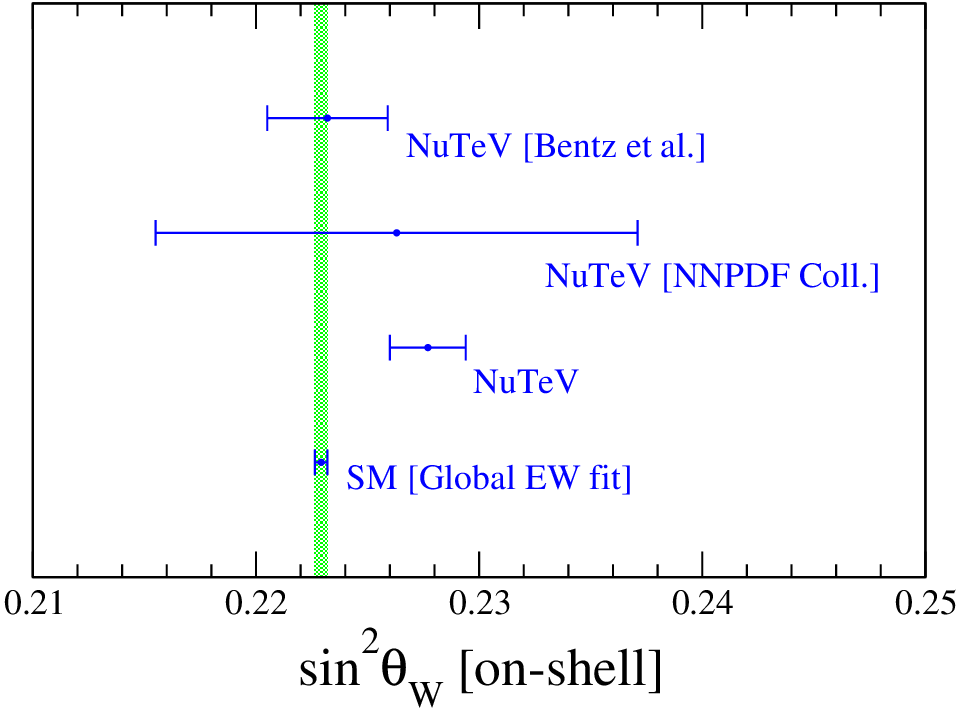}
\caption{Current status of $\sin^2\theta_W$ determinations from
  Refs.~\cite{Nakamura:2010zzi,Zeller:2001hh,Ball:2009mk,Bentz:2009yy}.}
\label{fig:sin2}
\end{center}
\end{figure*}

Ascribing the discrepancy between the NuTeV result for the left
coupling $(g^L_\mu)^2$ and its prediction within the Standard Model to
the existence of a non-zero NSI-operator, the authors in
Ref.~\cite{Davidson:2003ha} obtained a positive hint for non-zero
values of the left flavor-conserving NSI couplings
$\eps^{dL}_{\mu\mu}$ and $\eps^{uL}_{\mu\mu}$. On the other hand, for
the case of flavor changing NSI parameters they obtained limits on the
NSI $|\eps_{\mu\tau}^{qL;R}|$ couplings.

However, several corrections and theoretical uncertainties coming, for
instance, from nuclear effects and next-to-leading-order corrections
were neglected in the original NuTeV Collaboration analysis.
Although the presence of NSI could explain the discrepancy of the
NuTeV results with the SM, before claiming that they provide a hint of
new physics, one needs make sure that all uncertainties are carefully
taken into account.
As a matter of fact several attempts have been made to interpret the
NuTeV results just in terms of conventional physics, see for example
Refs.\cite{Davidson:2001ji,McFarland:2003jw}.

Here we re-analyze the impact of taking into account carefully the
 uncertainties in deriving restrictions on physics beyond the
SM as parametrized by the NSI Lagrangian in Eq.~(\ref{eq:efflag}).
In particular, we will focus on the two most recent re-analysis of the
NuTeV data given in Refs.~\cite{Ball:2009mk,Bentz:2009yy} in order to
obtain constraints on neutrino NSI coupling strengths.

The NNPDF Collaboration reports more precise estimates of the strange
and anti-strange parton distribution functions, leading to a new value
for the electroweak mixing angle~\cite{Ball:2009mk}:
\begin{equation}
\sin^2\theta_W = 0.2263 \pm 0.0014^{stat}\pm 0.0009^{sys}\pm 0.0107^{PDFs}.
\label{sin:nnpdf}
\end{equation}

On the other hand the analysis in Ref.~\cite{Bentz:2009yy} takes into
account three different corrections coming from nuclear effects, due
to the excess of neutrons in iron, charge symmetry violation arising
from up and down quark mass differences, and strange quarks.  With all
these corrections included, the following value for the electroweak
mixing angle is extracted:
\begin{equation}
\sin^2\theta_W = 0.2232 \pm 0.0013^{stat}\pm 0.0024^{sys}
\label{sin:bentz}
\end{equation}

\begin{figure*}
\begin{center}
\includegraphics[width=.45\textwidth,angle=0]{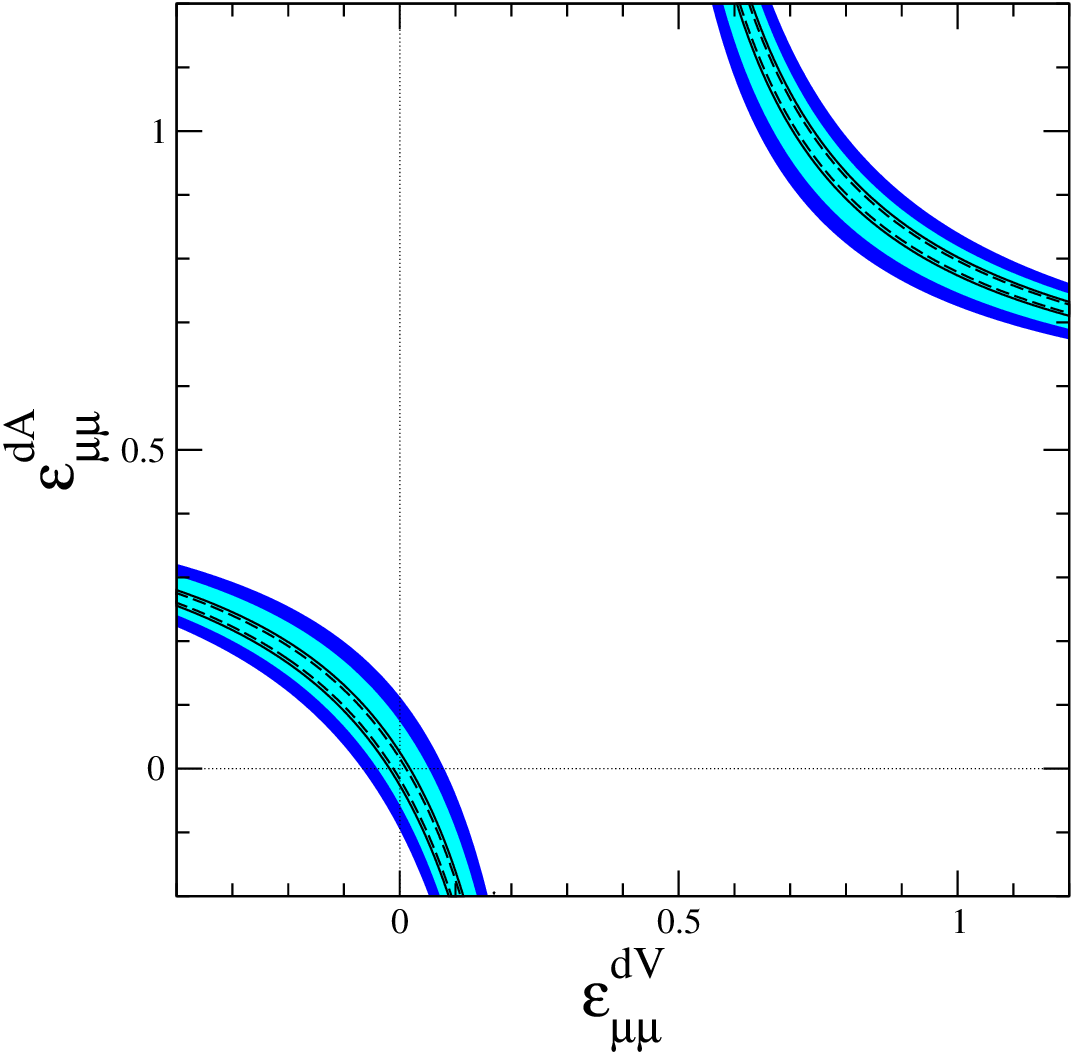}
\includegraphics[width=.45\textwidth,angle=0]{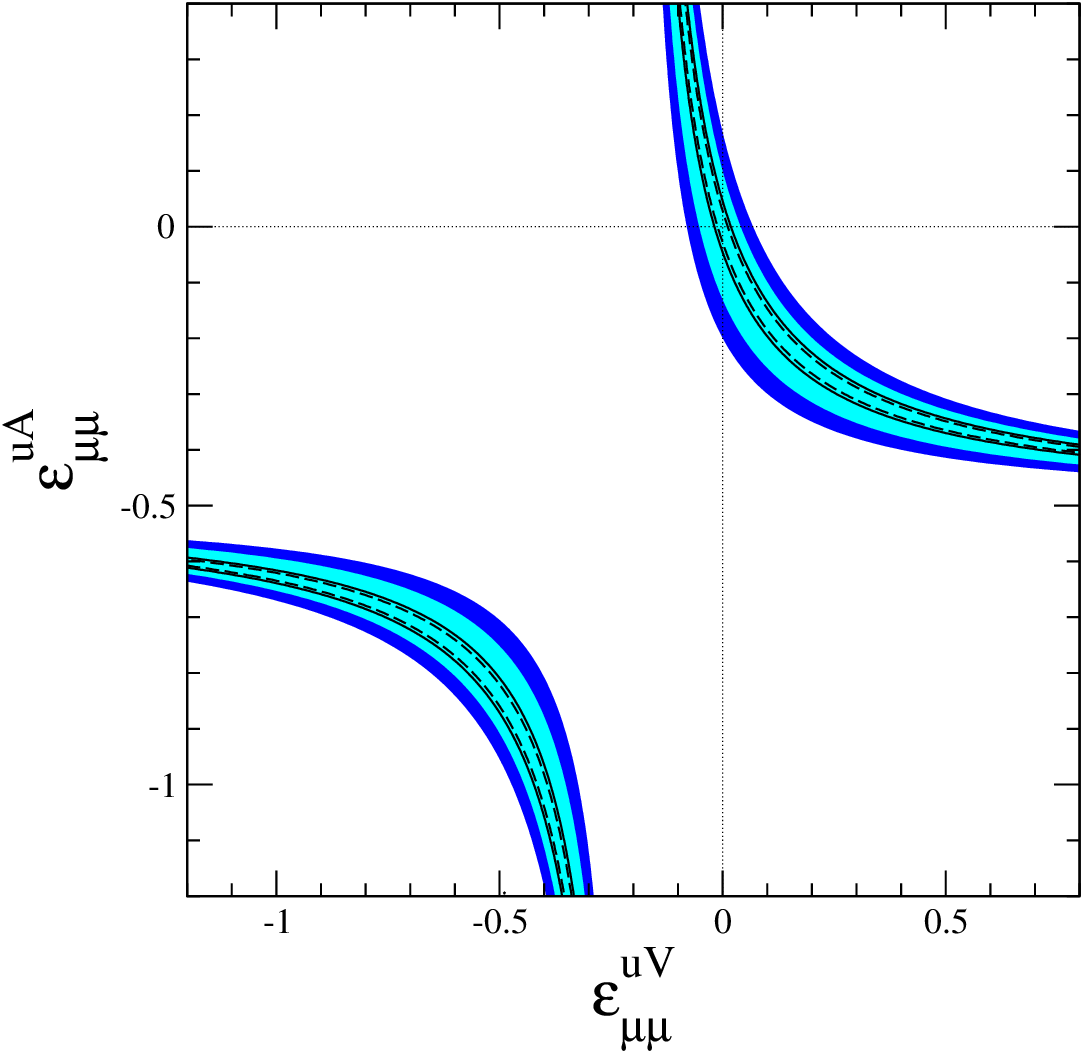}
\caption{Allowed regions at 90\% C. L. and 3$\sigma$ for the vectorial
  and axial NSI couplings of neutrinos with down (left panel) and
  up-type (right panel) quarks. Colored regions are obtained using
  NNPDF corrections~\cite{Ball:2009mk}, while empty lines correspond
  to Bentz et al~\cite{Bentz:2009yy}.}
\label{fig:nsi-fd}
\end{center}
\end{figure*}

One can see from Eqs. (\ref{sin:nnpdf}) and (\ref{sin:bentz}) that the
two re-calculated estimates for the electroweak mixing angle, with
larger uncertainties, are consistent with the SM prediction, as seen
in Fig.~\ref{fig:sin2}.
In order to determine the resulting restrictions set by NuTeV data on
the strength of NSI couplings, in view of the corrections discussed
above~\cite{Ball:2009mk,Bentz:2009yy}, we adopt the Paschos-Wolfenstein
ratio $R_{PW}$.
We perform a $\chi^2$ analysis using the Paschos-Wolfenstein ratio
derived from each re-analysis of NuTeV data, and we compare these
results with the Standard Model prediction for $R_{PW}$. As a result
of this simple statistical analysis, and allowing for one non-zero NSI
coupling at a time, we obtain the following constraints at 90\% C.L. :
\begin{eqnarray}
\label{eq:bounds-LR1}
& -0.017 < \eps^{dL}_{\mu\mu} < 0.025 \quad \& \quad 0.84 < \eps^{dL}_{\mu\mu} < 0.88 , &
\qquad -0.24 < \eps^{dR}_{\mu\mu} < 0.088 .
\\
& -0.72< \eps^{uL}_{\mu\mu} < -0.67 \quad \& \quad -0.031< \eps^{uL}_{\mu\mu} < 0.020 , &
\qquad  -0.058 < \eps^{uR}_{\mu\mu} < 0.063 \quad \& \quad 0.24 <
\eps^{uR}_{\mu\mu} < 0.36.
\label{eq:bounds-LR2}
\end{eqnarray}
when using the results in NNPDF, and
\begin{eqnarray}
\label{eq:bounds-LR3}
& -0.005 < \eps^{dL}_{\mu\mu} < 0.005 \quad \& \quad 0.86 < \eps^{dL}_{\mu\mu} < 0.87 , &
\qquad -0.17 < \eps^{dR}_{\mu\mu} < -0.11 \quad \& \quad -0.042 < \eps^{dR}_{\mu\mu} < 0.025 . 
\\
&-0.71< \eps^{uL}_{\mu\mu} <0.70 \quad \& \quad -0.006 < \eps^{uL}_{\mu\mu} < 0.006  , &
\qquad  -0.014 < \eps^{uR}_{\mu\mu} <0.016 \quad \& \quad 0.28 <
\eps^{uR}_{\mu\mu} < 0.31 .
\label{eq:bounds-LR4}
\end{eqnarray}
for the Bentz et al. case. For both analyses we observe two
allowed regions for most of the NSI couplings, reflecting the
degeneracy seen in Fig.~\ref{fig:nsi-fd}.

\begin{figure*}
\begin{center}
  \includegraphics[width=.5\textwidth,angle=0]{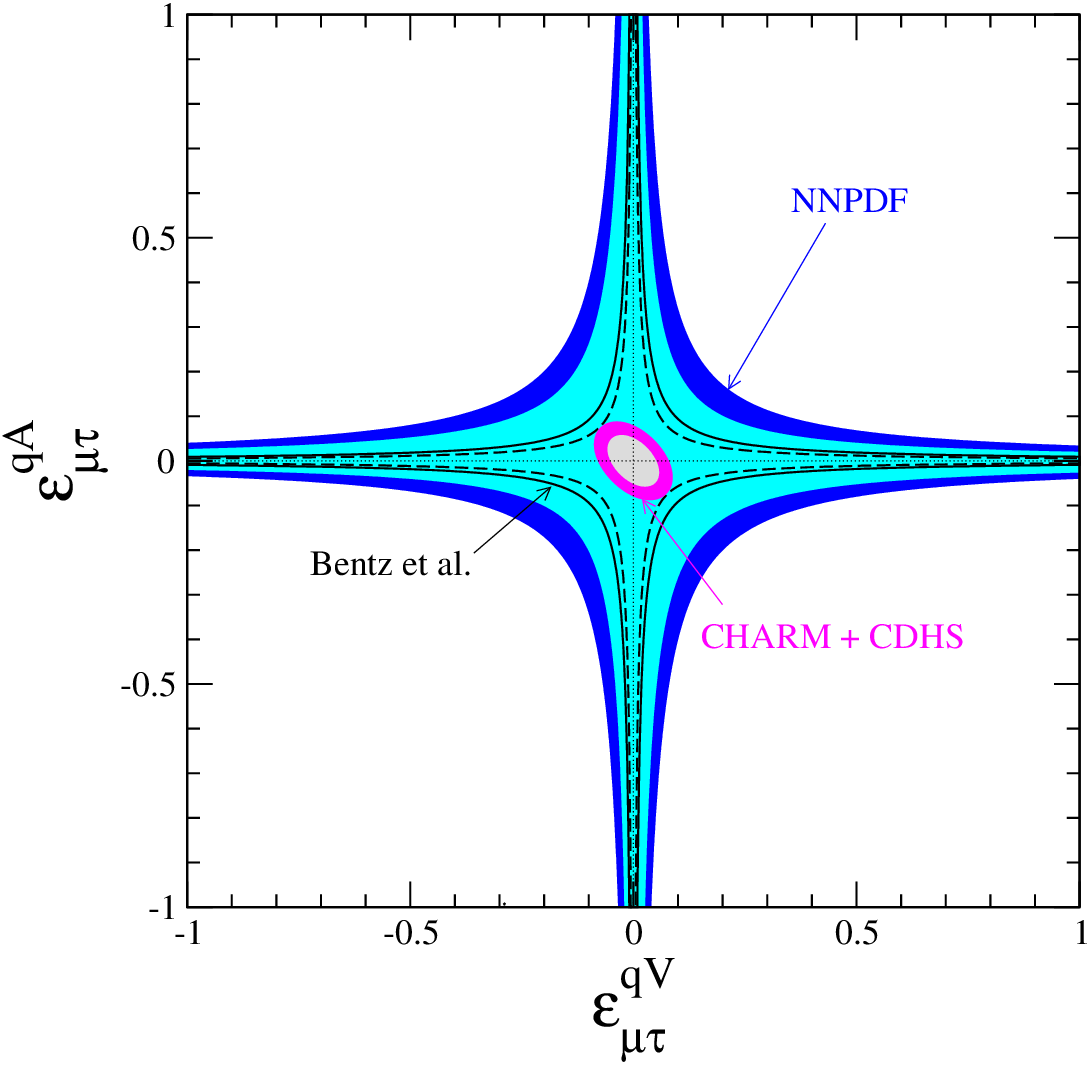}
  \caption{Same as Fig. \ref{fig:nsi-fd} for the flavor changing
    couplings of up- and down-type quarks with neutrinos. Colored
    regions are obtained using NNPDF corrections~\cite{Ball:2009mk},
    while empty lines correspond to the Bentz et
    al. re-analysis~\cite{Bentz:2009yy}. The small region in the colored ellipse
    results from the analysis of CHARM and CDHS results, see
    discussion in Secs.~\ref{sec:constr-from-charm} and
    \ref{sec:concl-future-prosp}.}
\label{fig:nsi-fc}
\end{center}
\end{figure*}

Allowing now for the simultaneous presence of left and right--handed
NSI neutrino couplings we obtain the results given in
Fig.~\ref{fig:nsi-fd}. For convenience, we present the constraints in
terms of the vector and axial couplings instead of using the $L,R$
basis.  This makes it easier to proceed when combining with the
atmospheric data (see below). The left panel holds for the case of a
down-type quark, while the right one holds for an up-type quark.
One can see that, in both cases, there is a two-fold degeneracy in the
values of the NSI parameters, which is easily understood from
Eq.~(\ref{eq:PW})\footnote{Note that in terms of the NSI couplings
  $\eps_{\mu\mu}^{qL}$, $\eps_{\mu\mu}^{qR}$, this equation represents
  a hyperbola centered at ($\eps_{\mu\mu}^{qL}$,
  $\eps_{\mu\mu}^{qR}$) = (-$g_\mu^{qL}$, -$g_\mu^{qR}$). After the
  change of variables to ($\eps_{\mu\mu}^{qV}$, $\eps_{\mu\mu}^{qA}$),
  the hyperbola is rotated to the one shown in Fig.~\ref{fig:nsi-fd}.
}.
On the other hand, for the case of flavor changing NSI, our analysis
gives the allowed region shown in Fig.~\ref{fig:nsi-fc}. In this case
the allowed regions for the two chiral components of the
flavor-changing NSI couplings of $u$- and $d$-type quarks are the
same, hence we show our results in only one plot. Note that here the
allowed region is a hyperbola centered at the origin
($\eps_{\mu\tau}^{qV}$, $\eps_{\mu\tau}^{qA}$) = (0,0), displaying a
remnant of the two-fold degeneracy seen for the flavor conserving
case.
In agreement with Eqs.~(\ref{eq:gl-nsi})-(\ref{eq:PW}), our analysis
constrains the product $\eps_{\mu\tau}^{qV} \cdot
\eps_{\mu\tau}^{qA}$, and therefore, one coupling could be of order
one, provided that the other one is small enough. As before, the
different widths for the allowed regions for each re-analysis of NuTeV
data reflect the experimental errors calculated in each case.

\section{Constraints from CHARM and CDHS}
\label{sec:constr-from-charm}

Both the CHARM and CDHS experiments have measured semileptonic
neutrino scattering cross sections. Although these early experiments
have a lower sensitivity than that reached at the NuTeV experiment,
their data are still useful in our global analysis.  While these two
experiments can not compete with the NuTeV sensitivity in constraining
the strength of non-universal NSI couplings, they play an important
role in restricting that of the flavor changing NSI~\footnote{The only
  caveat, however, is the bad quality of the resulting fit, given that
  these old experiments are not in good agreement with the SM, and
  that the addition of NSI only makes it worse.}  and for this reason
we consider them in our analysis. The measurements of these
experiments are summarized in Table~\ref{table:exp-values}.
\begin{table}
 \begin{tabular}{lccc}
\hline \hline Experiment & Observable & measurement & SM prediction  \\ \hline\hline
CDHS~\cite{Blondel:1989ev} & $R^\nu$ & $0.3072\pm 0.0033$ & $0.3208$
\\ & $R^{\bar{\nu}}$ & $0.382\pm 0.016$ & $0.381$ \\ & $r$ & $0.393\pm 0.014$ & \\ 
\hline
CHARM~\cite{Allaby:1987vr} & $R^\nu$ & $0.3093\pm 0.0031$ & $0.3226$\\ &
$R^{\bar{\nu}}$ & $0.390\pm 0.014$ & $0.371$\\ & $r$ & $0.456\pm 0.011$ &\\ \hline\hline
\end{tabular}
\caption{Ratios of neutral to charged currents measured by CHARM and CDHS compared with the SM prediction.}
\label{table:exp-values}
\end{table}
Using the information given in this table we perform a $\chi^2$
analysis, including the correlation between the neutrino and
antineutrino ratios of neutral to charged currents given by the
parameter $r$. Our $\chi^2$ function in this case is given by
\begin{equation}
\chi^2 = \sum_i \chi^2_i = \sum_{j,k} (R^{j}_{i}-R^{j}_{i,NSI}) 
         (\sigma^2)^{-1}_{jk}(R^{k}_{i}-R^{k}_{i,NSI})
\end{equation}
where $i$ runs for either CHARM or CDHS and $j,k$ stands for $R^{\nu}$
and $R^{\bar{\nu}}$. The results of our computation are shown in
Fig.~\ref{fig:nsi-fc}. Although the inclusion of these data strongly
constrains the axial flavor changing NSI parameters, the restriction
follows from a discrepancy between the experimental and the
theoretical value of $R^\nu$, as can be seen from Table
\ref{table:exp-values}. As already mentioned, this results in a very a
poor fit, hence we warn that constraints obtained for this case should
be considered as less robust than those obtained for non-universal NSI
discussed above. For the sake of completeness, however, and in analogy
with Eqs.(\ref{eq:bounds-LR1})-(\ref{eq:bounds-LR4}) we quote the
constraints that would be obtained at 90\% C.L. by combining
CHARM/CDHS with the re-analysis of NuTeV data performed by the NNPDF
Collaboration (or Bentz et al.):
\begin{equation}
 -0.023 (-0.023) < \eps^{qL}_{\mu\tau} < 0.023 (0.023), 
\qquad -0.039 (-0.036) < \eps^{qR}_{\mu\tau} < 0.039 (0.036) . 
\end{equation}

\section{Combining with atmospheric data}
\label{sec:comb-with-atmosph}

Atmospheric neutrino data are well described by the standard mechanism
of neutrino oscillations~\cite{schwetz:2008er,maltoni:2004ei}. As a
result, if neutrino NSI with matter exist, they must at best play a
sub-leading role in the description of atmospheric neutrino data.
We now discuss atmospheric neutrino propagation in the presence of
non--standard neutrino--matter interactions, containing both
flavor-changing and flavor-diagonal
components~\cite{fornengo:2001pm,gonzalezgarcia:2007ib,Mitsuka:2008zzb}.
In the context of such hybrid scheme, the presence of neutrino NSI
would affect atmospheric neutrino propagation in matter introducing an
extra term in the evolution Hamiltonian.
In the simple two--neutrino description the NSI contribution to the
Hamiltonian will be given by
\begin{equation}
H_{NSI} = \sqrt{2} G_F N_q \left(
\begin{array}{cc}
\eps_{\mu\mu}^{qV}  & \eps_{\mu\tau}^{qV} \\
\eps_{\mu\tau}^{qV}  & \eps_{\tau\tau}^{qV} \\
\end{array}
\right)
\end{equation}
with q=u, d and $\eps_{\alpha\beta}^{qV} = \eps_{\alpha\beta}^{qL} +
\eps_{\alpha\beta}^{qR}$, where one notes that neutrino propagation is
only sensitive to the vectorial NSI couplings. This provides important
information complementary to that coming from the accelerator
experiments, see Eqs.~(\ref{eq:gl-nsi}) and (\ref{eq:gr-nsi}).

The Super-Kamiokande neutrino data have been analyzed under this
assumption in
Refs.~\cite{fornengo:2001pm,gonzalezgarcia:2007ib,Mitsuka:2008zzb}. Up
to now no evidence of NSI has been found in the atmospheric data
sample and, as a result, one gets upper bounds on the magnitude of the
NSI coupling strengths.
Here we will include in our analysis the most recent published results
using the full atmospheric SK-I and SK-II data
sample~\cite{Mitsuka:2008zzb} which leads to the following bounds:
\begin{eqnarray}
 0.007 < & \eps_{\mu\tau}^{dV}  & <   0.007 \\ 
& | \eps_{\tau\tau}^{dV} - \eps_{\mu\mu}^{dV}|  & <  0.042~,
\label{eq:bounds-atm-d}
\end{eqnarray}
at 90\% C.L. (1 d.o.f.). These limits have been obtained for the case
of neutrino NSI with down-type quarks. Taking into account the
chemical composition of the Earth in the PREM
model~\cite{Dziewonski:1981xy}, we can rewrite the bounds on
Eq.~(\ref{eq:bounds-atm-d}) in terms of the neutrino NSI couplings
with up-type quarks to a good approximation. According to
Ref.~\cite{Lisi:1997yc}, the ratio between the density of down and up
quarks $N_d/N_u$ takes a value of 1.009 (1.047) in the mantle (core)
of the Earth.  Using the average value: $N_d/N_u = 1.028$ one obtains
the following 90\% C.L. bounds on the neutrino - up quark NSI
couplings:
\begin{eqnarray}
 0.007 < & \eps_{\mu\tau}^{uV}  & <   0.007 \\ 
& | \eps_{\tau\tau}^{uV} - \eps_{\mu\mu}^{uV}|  & <  0.043
\label{eq:bounds-atm-u}
\end{eqnarray}

\begin{figure*}
\begin{center}
\includegraphics[width=.45\textwidth,angle=0]{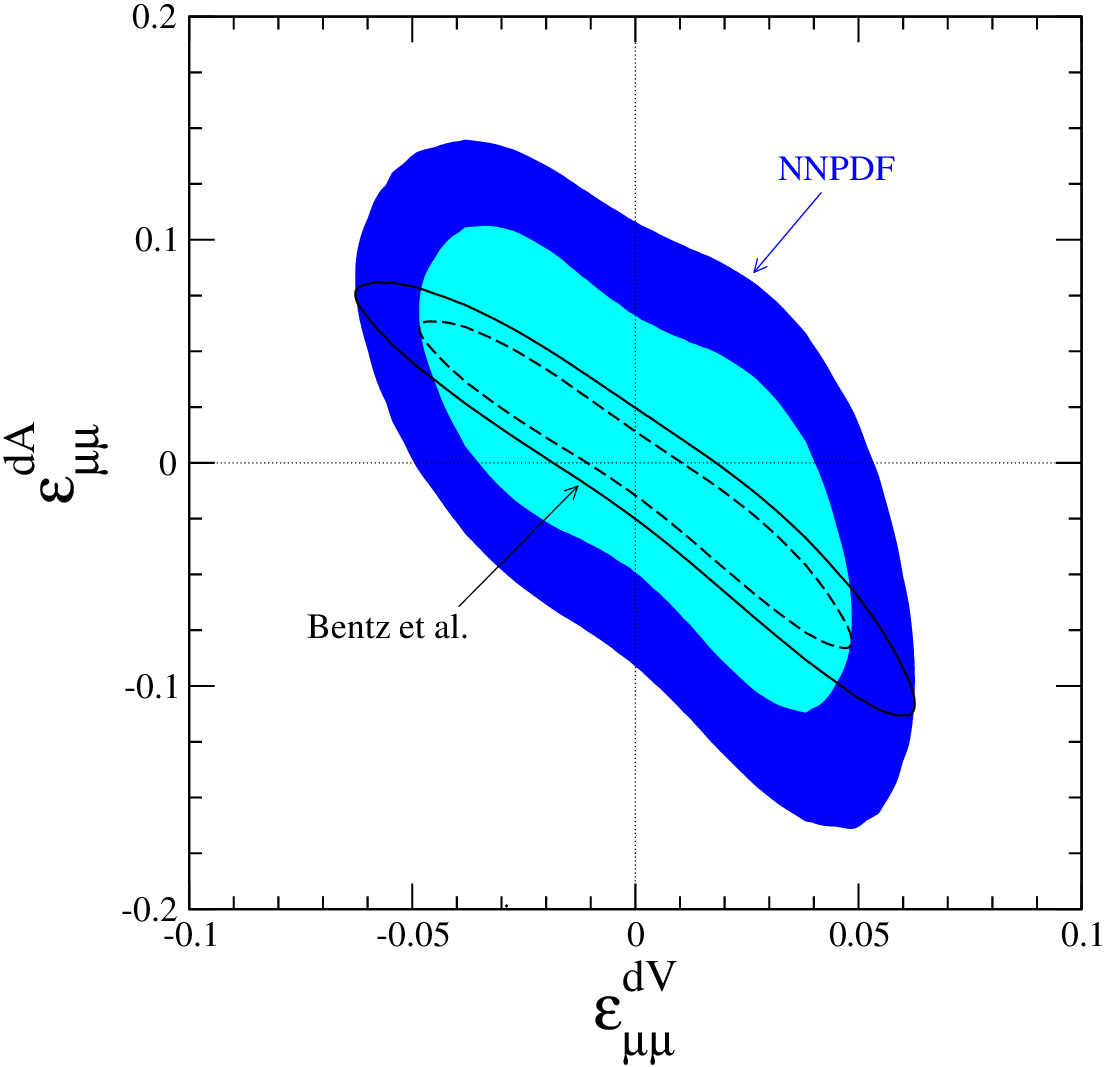}
\includegraphics[width=.45\textwidth,angle=0]{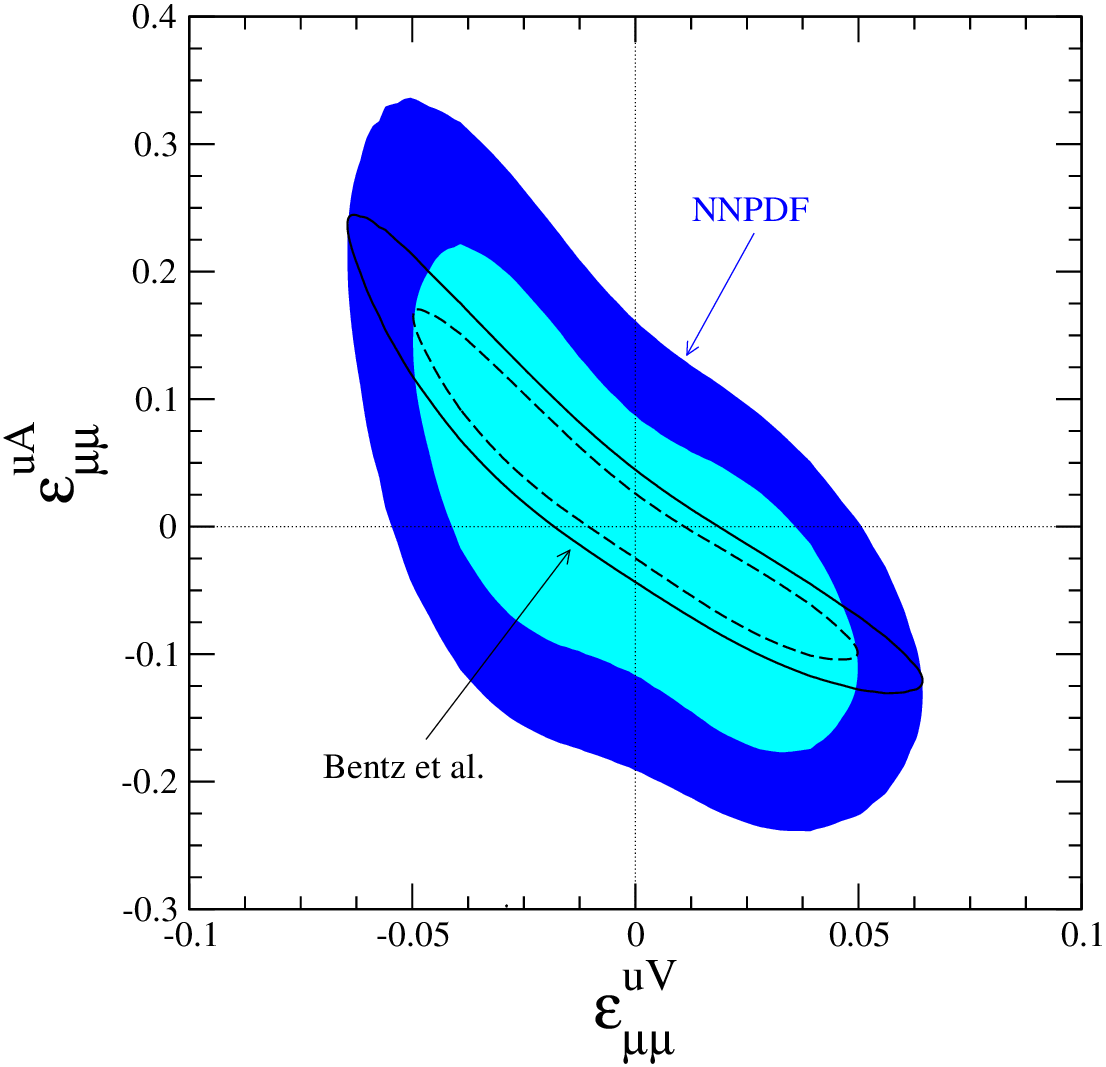}
\caption{Allowed regions at 90\% C. L. and 3$\sigma$ for
  flavor-diagonal NSI couplings of down- (left panel) and up-type
  quarks (right panel) with neutrinos obtained by combining
  accelerator data with results from the atmospheric data analysis. 
  We use the same conventions as in Fig 2.}
\label{fig:nsi-fd-atm}
\end{center}
\end{figure*}

\begin{figure*}
\begin{center}
  \includegraphics[width=.45\textwidth,angle=0]{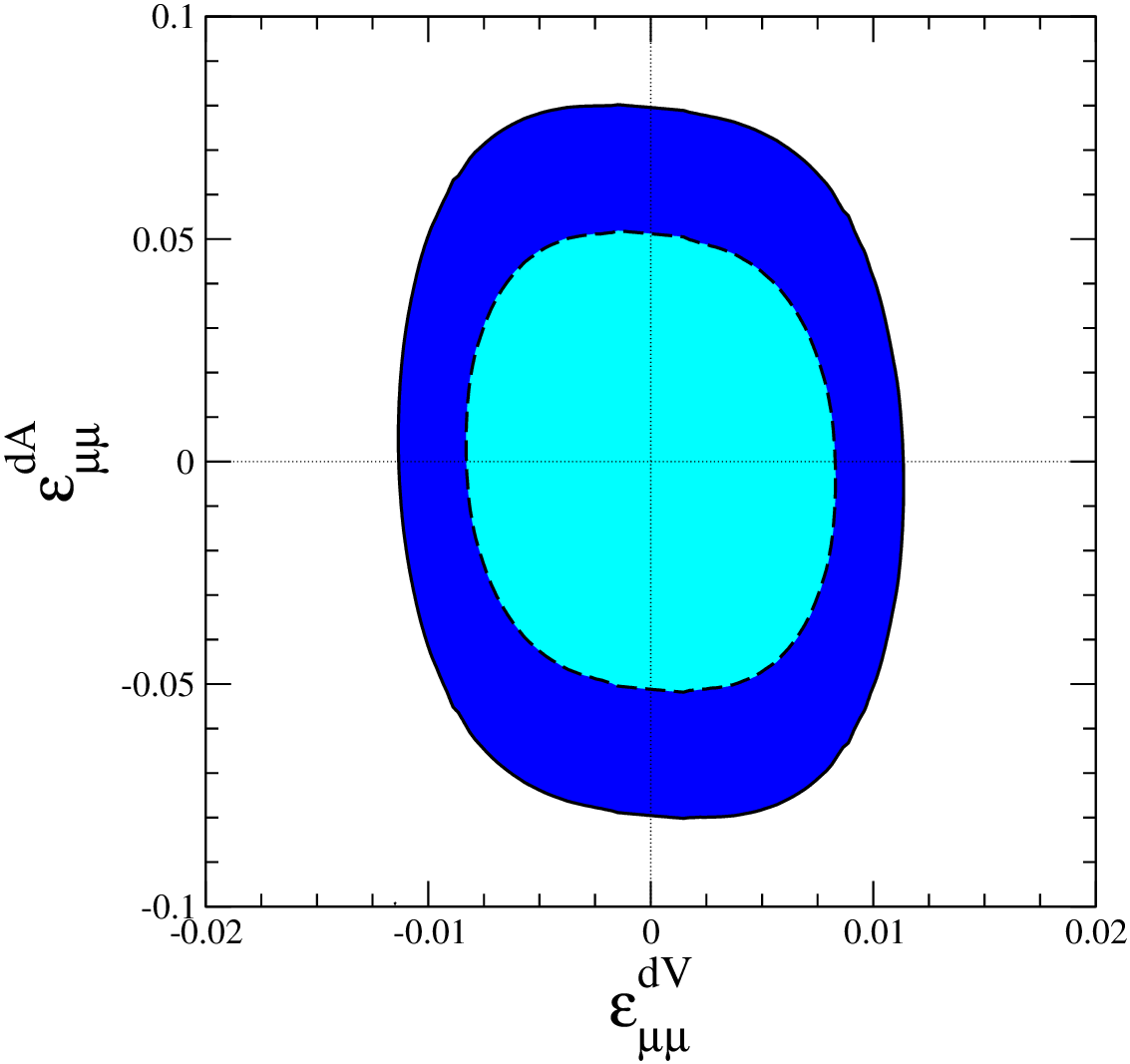}
 \caption{Same as Fig. \ref{fig:nsi-fd-atm} for the flavor changing
   NSI couplings with down-type quarks, with nearly the same results
   for up-type quarks. The two re-analyses on NuTeV data considered
   here lead to the same results, since the analysis is dominated by
   atmospheric and CHARM + CDHS data.}.
\label{fig:nsi-fc-atm}
\end{center}
\end{figure*}

We now go one step further combining also these results with those of
the previous accelerator analysis.
The constraints derived from such a combined analysis are shown in
Figs. \ref{fig:nsi-fd-atm} and \ref{fig:nsi-fc-atm} for the case of
flavor-diagonal and flavor-changing NSI, respectively. 
As before, for the non-universal NSI couplings we consider NuTeV,
while for the flavor-changing couplings we also include the results of
CHARM and CDHS.
As seen in the previous section, for the case of flavor changing NSI
the NuTeV accelerator results and atmospheric neutrino bounds are
nearly the same for up- or down-type quarks, and therefore we only
present the plot for down-type quarks.
One sees that the atmospheric neutrino data provide quite a useful
tool to constrain the vector NSI coupling parameter, allowing to
remove the degeneracy discussed before.

Once we make the projection of the two parameter analysis into one
parameter, we obtain the following constraints at 90\% C.L. when
considering the reanalysis of the NuTeV anomaly reported by
Ref.~\cite{Ball:2009mk} (\cite{Bentz:2009yy}):
\begin{eqnarray}
& -0.042 (-0.042) < \eps^{dV}_{\mu\mu} < 0.042 (0.042), &
\qquad -0.091 (-0.072) < \eps^{dA}_{\mu\mu} < 0.091 (0.057) . 
\\
& -0.044 (-0.044)< \eps^{uV}_{\mu\mu} < -0.044 (0.044), &
\qquad  -0.15 (-0.094)< \eps^{uA}_{\mu\mu} < 0.18 (0.14).
\end{eqnarray}

For the case of the flavor changing parameters the results are nearly
the same for the two different re-analysis of NuTeV data, since the
determination of the vectorial couplings $\eps_{\mu\tau}^{qV}$ are
dominated by the atmospheric neutrino analysis, while the axial
couplings $\eps_{\mu\tau}^{qA}$ are mainly given by the accelerator
experiments CHARM and CDHS. One finds,
\begin{eqnarray}
& -0.007 < \eps^{dV}_{\mu \tau} < 0.007, &   
\qquad -0.039 < \eps^{dA}_{\mu \tau} < 0.039 .
\\
& -0.007 < \eps^{uV}_{\mu \tau} < 0.007, &   
\qquad  -0.039 < \eps^{uA}_{\mu \tau} < 0.039 .
\end{eqnarray}
A more complete full-fledged three-neutrino global analysis in the
context of neutrino oscillations plus non-standard interactions will
introduce three additional NSI couplings, namely $\eps_{ee}$,
$\eps_{e\mu}$ and $\eps_{e\tau}$ and, as a result, the bounds would be
correspondingly weaker.

\begin{table}
 \begin{tabular}{lccc}
\hline \hline Global with NuTeV reanalysis & NSI with down & NSI with up  \\
\hline \hline
 & NU & NU\\ \hline
 NNPDF~\cite{Ball:2009mk}  
   & $-0.042 < \eps^{dV}_{\mu\mu} < 0.042$ &
$-0.044< \eps^{uV}_{\mu\mu} < -0.044$ &
\\ & $-0.091 < \eps^{dA}_{\mu\mu} < 0.091$ & $-0.15 <
\eps^{uA}_{\mu\mu} < 0.18$ \\
\hline
Bentz at al.~\cite{Bentz:2009yy} 
 & $-0.042 < \eps^{dV}_{\mu\mu} <
0.042$ & $-0.044< \eps^{uV}_{\mu\mu} < -0.044$ &\\ & $-0.072 <
\eps^{dA}_{\mu\mu} < 0.057$ & $-0.094 < \eps^{uA}_{\mu\mu} < 0.14$ \\
\hline  \hline 
& FC & FC \\
\hline 
NNPDF/Bentz et al.
 & $-0.007 < \eps^{dV}_{\mu \tau} < 0.007$ & $-0.007 < \eps^{uV}_{\mu \tau} < 0.007$ & \\
& $-0.039 < \eps^{dA}_{\mu \tau} < 0.039$ & $-0.039 < \eps^{uA}_{\mu \tau} < 0.039$ & \\  
\hline \hline 
\end{tabular}
\caption{Allowed values for NSI parameters from a global analysis of NuTeV 
  data using NNPDF and Bentz et al. corrections, combined with atmospheric
  data. The top of the table corresponds to  non-universal (NU) NSI 
  parameters and the bottom to flavor-changing (FC) NSI. In the latter 
  case CDHS and CHARM data are also included.}
\label{table:final results}
\end{table}

\section{Discussion and future prospects}
\label{sec:concl-future-prosp}

We have re-analysed the constraints on novel non-standard neutrino
interactions of muon neutrinos with quarks which follow from current
accelerator experiments. In particular, we have re-considered the
results of the NuTeV experiment in view of a new evaluation of the
NuTeV systematic uncertainties. We have combined the restrictions
following from accelerator data with those coming from recent
atmospheric data analysis, which plays a crucial role in removing
degeneracies. We have found that, although constraints for muon
neutrinos are better than those applicable to tau or electron
neutrinos, they are not as strong as previously believed. While
previous authors report bounds of the order of few $\times 10^{-3}$,
we find that, taking into account the current estimates of the NuTeV
uncertainties, constraints are now of the order of few $\times
10^{-2}$. Our results are summarized in Table~\ref{table:final
  results}. Even weaker constraints would hold in a generalized
three-generation framework. Although muon neutrino NSI couplings are
often neglected~\cite{Escrihuela:2009up}, one should keep in mind the
weakness of the existing limits.

Note that these results will not be improved by the inclusion of data
from the long baseline experiment MINOS \cite{Michael:2006rx}, given
its poor sensitivity to matter effects when compared with that of
atmospheric data, at least in a two-neutrino analysis. As a result
MINOS will not significantly restrict the neutral current NSI
couplings discussed here, and the same is expected for
OPERA~\cite{Blennow:2008ym}.
%
%
However, future "clean" measurements of the oscillation parameters
combined with the NSI-contaminated parameters determined by
atmospheric experiments would be helpful.
Certainly it would also help considerably to have a very long baseline
setup as envisaged in the Long Baseline Neutrino Experiment (LBNE)
project currently proposed, capable of probing Earth matter effects
with enhanced sensitivity~\cite{diwan}.

Last, but not least, we note that a new high statistics neutrino
scattering experiment, NuSOnG has been proposed~\cite{Adams:2008cm},
with the same goals as NuTeV. One of its main tasks would be to probe
new physics in the neutrino couplings.
Expectations are that NuSOnG would improve the NuTeV sensitivities on
muon-neutrino quark scattering by a factor 2 at least, which would
translate into a similar improvement on the NSI sensitivities with
respect to those obtained from NuTeV.

\begin{acknowledgments}

We would like to thank I. Cloet for useful comments about his work in Ref.~\cite{Bentz:2009yy}.
Work supported by MICINN grants FPA2008-00319/FPA, CSD2009-00064, by
PROMETEO/2009/091, by CONACyT and by EU network
PITN-GA-2009-237920. M.T.\ acknowledges financial support from CSIC
under the JAE-Doc programme.

\end{acknowledgments}


\end{document}